\documentclass[preprint,nofootinbib,12pt,onecolumn]{revtex4}
\pdfoutput=1

\usepackage{setspace}
\usepackage[letterpaper,margin=1in]{geometry}

\usepackage{epsfig}
\usepackage{amssymb}

\usepackage{graphicx}
\usepackage{dcolumn}
\usepackage{bm}
\usepackage{subfigure}
\usepackage{epsfig}
\usepackage{amsmath}
\usepackage{amsfonts}
\usepackage{graphicx}
\usepackage{amssymb}
\usepackage{nccmath}
\usepackage{amssymb,amsmath,amsfonts}
\usepackage{xfrac}
\usepackage{color}
\usepackage{tikz}
\usepackage[T1]{fontenc}
\usepackage[utf8]{inputenc}
\usepackage{ulem}

\usepackage{tcolorbox}
\usepackage{hyperref}

\let\oldsqrt\sqrt
\def\sqrt{\mathpalette\DHLhksqrt}
\def\DHLhksqrt#1#2{%
\setbox0=\hbox{$#1\oldsqrt{#2\,}$}\dimen0=\ht0
\advance\dimen0-0.2\ht0
\setbox2=\hbox{\vrule height\ht0 depth -\dimen0}%
{\box0\lower0.4pt\box2}}

\newcommand{\sss}[1]{{\scriptscriptstyle{#1}}}

\newcommand{\uPl}{\mathrm{Pl}}

\newcommand{\usssPl}{\sss{\uPl}}

\newcommand{\Mp}{M_\usssPl}

\newcommand{\beq}{\begin{equation}}
\newcommand{\eeq}{\end{equation}}
\newcommand{\bea}{\begin{equation}\begin{aligned}}
\newcommand{\eea}{\end{aligned}\end{equation}}

\newlength{\wsingfig}
\newlength{\wdblefig}
\newlength{\wquadfig}
\newlength{\wtriplefig}
\newcommand{\Eq}[1]{Eq.~(\ref{#1})}

\newcommand{\Fig}[1]{Fig.~{\ref{#1}}}

\begin{document}

\begin{titlepage}

\begin{center}

{\Large \bf  
Primordial black hole driven cosmic acceleration
}

\vspace{.5cm}
\bf Konstantinos Dialektopoulos $^{a,b\ast}$, Theodoros Papanikolaou
$^{c,d,\dagger}$,   \\Vasilios Zarikas $^{e,\ddagger}$
\end{center}

{\small
\begin{quote}
\begin{center}
$^a$ Department of Mathematics and Computer Science,
Transilvania University of Brasov, 500091, Brasov, Romania\\
$^b$ Institute of Space Sciences and Astronomy, University of Malta, Msida, Malta and Department of Physics, University of Malta, Msida, Malta\\
\texttt{kdialekt@gmail.com} \\
$^c$ Scuola Superiore Meridionale, Largo San Marcellino 10, 80138 Napoli, Italy\\
$^d$
Istituto Nazionale di Fisica Nucleare (INFN), Sezione di Napoli, Via Cinthia 21, 80126 Napoli, Italy\\
\texttt{t.papanikolaou@ssmeridionale.it}\\
$^e$ Department of Mathematics, University of Thessaly, 35100, Lamia, Greece\\
\texttt{vzarikas@uth.gr}\\
\end{center}%
\end{quote}}


\vspace{.1cm}
\begin{center}
    \small{Essay written for the Gravity Research Foundation 2025 Awards for Essays on Gravitation.}
\end{center} 
\vspace{.5cm}

\centerline{\bf Abstract}
We propose a natural mechanism for cosmic acceleration driven by primordial black holes (PBHs) with repulsive behavior, within a “Swiss Cheese” cosmological framework. Considering regular black hole spacetimes such as Hayward, Bardeen, and Dymnikova—as well as the singular Schwarzschild–de Sitter case—we consistently find a robust PBH-driven cosmic acceleration phase. This phase ends either at an energy scale set by the PBH parameters or through black hole evaporation. Notably, one finds that ultra-light PBHs with \( m < 5 \times 10^8 \) g can trigger exponential inflation with graceful exit and reheating. Additionally, PBHs with \( m \sim 10^{12} \) g and abundances \( 0.107 < \Omega^\mathrm{eq}_\mathrm{PBH} < 0.5 \) near matter-radiation equality can act as an early dark energy component, offering a potential resolution to the Hubble tension.

\vspace{0.5cm}
Corresponding Authors: K. Dialektopoulos, T. Papanikolaou and V. Zarikas
\end{titlepage}

\medskip
{\bf{Introduction}}--Primordial black holes (PBHs), first proposed in the 1970s~\cite{1967SvA....10..602Z, Carr:1974nx,1975ApJ...201....1C,1979A&A....80..104N}, led to the pivotal discovery of black hole evaporation by Hawking~\cite{Hawking:1975vcx}. PBHs can form through various mechanisms, such as the collapse of enhanced primordial perturbations~\cite{Carr:1974nx,Carr:1975qj,Musco:2020jjb} [see~\cite{Garcia-Bellido:1996mdl,Yokoyama:1995ex,Garcia-Bellido:2017mdw} for inflationary models], cosmological phase transitions~\cite{Hawking:1982ga,Moss:1994iq,Jung:2021mku}, scalar field instabilities~\cite{Khlopov:1985fch}, topological defects~\cite{Hawking:1987bn,Polnarev:1988dh} and modified or quantum gravity scenarios~\cite{Barrow:1996jk,Kawai:2021edk,Papanikolaou:2023crz}.  

PBHs have recently attracted interest for their potential to explain part or all of dark matter~\cite{Chapline:1975ojl,Carr:2020xqk}, structure formation~\cite{Meszaros:1975ef,Afshordi:2003zb}, as well as certain black hole mergers observed by LIGO/VIRGO~\cite{Sasaki:2016jop,Franciolini:2021tla}. They may also play a key role in the processes of reheating~\cite{Lennon:2017tqq}, baryogenesis~\cite{Barrow:1990he, Baumann:2007yr, Aliferis:2014ofa, Aliferis:2020dxr} and primordial magnetic field generation~\cite{Safarzadeh:2017mdy,Araya:2020tds,Papanikolaou:2023nkx}. For relevant reviews in the field of PBHs, see~\cite{Carr:2020gox,Escriva:2022duf}.

Most studies in the PBH literature assume Schwarzschild or Kerr metrics for PBHs~\cite{Khlopov:2008qy,Carr:2016drx,Escriva:2022duf,LISACosmologyWorkingGroup:2023njw,Choudhury:2024aji}. In this work, we instead explore PBHs with repulsive behavior, typically associated with regular spacetimes~\cite{Luongo:2023aib}, which may simultaneously address the dark matter and singularity problems~\cite{Easson:2002tg,Dymnikova:2015yma,Pacheco:2018mvs,Arbey:2021mbl,Arbey:2022mcd,Banerjee:2024sao,Davies:2024ysj,Calza:2024fzo,Calza:2024xdh}. For reviews on regular black holes, the interested reader can look at~\cite{Ansoldi:2008jw,Nicolini:2008aj,Torres:2022twv,Lan:2023cvz}.

In particular, in this essay we present a recently proposed novel mechanism for cosmic acceleration associated to regular PBHs or more general to PBHs with repulsive behaviour~\cite{Dialektopoulos:2025mfz}. Interestingly enough, this mechanism can account for an early inflationary era or accelerate the Universe at later times. In order to describe the collective influence of a population of PBHs on cosmic expansion, a ``Swiss Cheese'' cosmological model is adopted, originally introduced by Einstein and Strauss back in $1945$~\cite{Einstein:1945id}, where one matches one black hole metric with a homogeneous, isotropic Universe. When Schwarzschild black holes are used, this naturally yields a dust-filled Universe. However, as we will demonstrate followingly, if one accounts for PBH spacetimes with a repulsive behavior, they are met with an accelerated cosmic expansion. It is worth mentioning that this new swiss model under consideration is a fair approximation of a black hole dominated with PBHs early Universe which is not true for the description of the late cosmology stage. 


Remarkably, this mechanism can produce naturally an inflationary phase ending at a scale determined by the PBH metric or due to black hole evaporation, offering an alternative to the standard scalar filed inflationary/reheating scenario~\cite{Garcia-Bellido:1996mdl,Hidalgo:2011fj,Suyama:2014vga,Lennon:2017tqq,Zagorac:2019ekv}. It may also give rise to an early dark energy (EDE) component, without invoking scalar fields, potentially easing the Hubble tension.

{\bf{Regular black holes}}--Regular black hole solutions naturally emerge in quantum gravity frameworks that resolve curvature singularities, either through continuous quantum spacetimes~\cite{Cadoni:2023nrm, Bonanno:2023rzk, dePaulaNetto:2023cjw} or discrete geometries~\cite{Singh:2009mz,Ashtekar:2023cod}. In both approaches, the theory must explain the emergence of classical spacetime. Black holes can then be described by effective classical metrics with high-curvature modifications, yielding non-singular cores~\cite{1968qtr..conf...87B,Frolov:1981mz,Roman:1983zza,Dymnikova:1992ux,Borde:1996df,Hayward:2005gi,Hossenfelder:2009fc,Bambi:2013gva,Bambi:2013caa,Hawking:2014tga,Frolov:2014jva,Bardeen:2014uaa,Haggard:2014rza,Barrau:2015uca,Haggard:2015iya}.


Below, we illustrate our cosmic acceleration mechanism using the Hayward metric.
The original Hayward metric initially proposed in $2006$ can be recast as~\cite{Hayward:2005gi}
\begin{equation}\label{eq:Hayward_metric}
ds^2=-F(R)\,dt^2+\frac{1}{F(R)}\,dR^2+R^2\,d\Omega^2 ~,
\end{equation}
where
\begin{equation}\label{hayward}
F(R)=1-\frac{2\,G_\mathrm{N} M(R)}{R}~,
\end{equation}
and
\begin{align}
M(R)=\frac{m\,R^3}{R^3+2\,G_\mathrm{N}\, m\,L^2}  \approx
\begin{cases}
m & (R  \gg G_\mathrm{N}^{1/3} m^{1/3} L^{2/3}) \\
R^3/(2G_\mathrm{N} L^2) & (R  \ll G_\mathrm{N}^{1/3} m^{1/3} L^{2/3})  \, .
\end{cases}
\end{align}
Here, $c=1$, and $m$ represents the black hole mass at asymptotic infinity.

This form can be justified by several physical assumptions~\cite{Frolov:2017rjz}: (i) quantum corrections are negligible at low curvature (${\cal R}\ll L^{-2}$), where $L$ is a fundamental length scale; (ii) the metric remains regular everywhere; and (iii) the limiting curvature conjecture holds: $|{\cal R}|\le C L^{-2}$, with $C$ being a dimensionless constant depending on a curvature invariant constructed by the Riemann tensor~\cite{1982ZhPmR..36..214M,Markov:1984ii,Polchinski:1989ae}.

The metric \eqref{eq:Hayward_metric} admits Killing horizons, fixed by the roots of
\begin{equation} \label{hor}
F(R) = 1-\frac{2\,G_\mathrm{N}\,m\,R^2}{R^3+2\,L^2\,G_\mathrm{N}\, m}=0 \ , 
\end{equation}
which depend on $L$ and $m$. Specifically, for $m>\frac{3\sqrt{3}L}{4G_\mathrm{N}}$, the metric \Eq{eq:Hayward_metric} exhibits inner and outer horizons at approximately $R \simeq L$ and $R \simeq 2\,G_\mathrm{N}\,m$, respectively. Penrose diagram analyses of the Hayward metric can be found in~\cite{Hayward:2005gi,DeLorenzo:2014pta}.

The Bardeen and Dymnikova regular spacetimes are discussed in detail in the Appendices of \cite{Dialektopoulos:2025mfz}. In addition, the singular but repulsive de Sitter–Schwarzschild spacetime on large scales is also studied, and it can be viewed as a limiting case of the McVittie metric~\cite{1933MNRAS..93..325M}. Notably, this latter  McVittie black hole metric has been argued recently to approximate the general relativistic PBH metric during the collapse phase~\cite{DeLuca:2020jug,Hutsi:2021vha,Hutsi:2021nvs}.

{\bf{Cosmic acceleration emergence from cosmological ``Swiss Cheese'' matching}}--
Having introduced our regular PBH metric, one can describe now the collective behaviour of a population of PBHs dominating the energy of the Universe by matching the Hayward metric (being static and spherically symmetric) with an exterior homogeneous and isotropic spacetime. The technical details of this procedure are outlined in the Appendix A of~\cite{Dialektopoulos:2025mfz} as well as in~\cite{Kofinas:2017gfv}. Two clarifications regarding the use of the ``Swiss Cheese'' cosmological approach are in order. First, we apply it to the early Universe, where an approximately homogeneous distribution of black holes is a valid assumption~\cite{Desjacques:2018wuu, Ali-Haimoud:2018dau, MoradinezhadDizgah:2019wjf,DeLuca:2022uvz}, unlike the present epoch. Second, concerns about its stability~\cite{Krasinski:1997yxj} are irrelevant here, since we are not modeling isolated large-scale structures (e.g., galaxies). Instead, we study early cosmic expansion in a Universe filled with many black holes, making this approach well justified.

For a single Hayward black hole matched to an FLRW background, the background cosmic expansion equations become:
\begin{equation}\label{eq:H_2_Hayward}
H^2=\frac{\dot{a}^2}{a^2}=\frac{2G_\mathrm{N} m}{R^3+2 G_\mathrm{N} m L^2}-\frac{k}{a^2}\,,  \quad \frac{\ddot{a}}{a}=\frac{G_\mathrm{N} m (4\,G_\mathrm{N} L^2 m-R^3)}{(2 G_\mathrm{N} L^2 m+R^3)^2}\,, 
\end{equation}
where $ R = a\,r_{\Sigma} $, $ a $ is the scale factor, $ H=\dot{a}/a $ is the Hubble parameter and $\,r_{\Sigma} $ is the comoving Shucking radius which emerges out of the ``Swiss Cheese" cosmological analysis~[See ~\cite{Dialektopoulos:2025mfz} for more details.].

Generalizing thus our analysis to a Universe filled with many black holes, we model the latter  as a system of spherical vacuoles, each containing a black hole. Having thus a Universe  dominated by population of homogeneously distributed black holes we identify the energy density in the exterior region of a black hole with its interior mass density, i.e. 
\[
\rho = \frac{3m}{4\pi(a r_{\Sigma})^3},
\]
Substituting then into the second Eq. of~\eqref{eq:H_2_Hayward}, the background cosmic acceleration equation becomes:
\begin{equation}
\frac{\ddot{a}}{a}= 4 \pi G_\mathrm{N} \rho \,  \frac{(16 G_\mathrm{N} L^2 \pi \rho - 3)}{(8 G_\mathrm{N} L^2 \pi \rho + 3)^2}.
\label{acceleration with density}
\end{equation}
The sign of $\ddot{a}/a$ depends on the term $16 G_\mathrm{N} L^2 \pi \rho - 3$, with the critical energy density for the transition from a cosmic acceleration to a cosmic deceleration phase given by
\begin{equation}\label{rhoch}
\rho_{\rm c}=\frac{3}{16 \pi G_\mathrm{N}L^2}.
\end{equation}
As one may infer from \Eq{acceleration with density}, for $\rho > \rho_{\rm c}$, the Universe undergoes an accelerated cosmic expansion, while $\rho < \rho_{\rm c}$ leads to a standard cosmic deceleration phase. Thus, inflation ends naturally, without any fine-tuning.


{\bf{What are the cosmological consequences of a PBH driven cosmic acceleration?}}-- Having deduced above the cosmic expansion dynamics of a Universe dominated by repulsive PBHs, we investigate here two key cosmological consequences of such a phenomenology at the levels of inflation and early dark energy.

\textit{Inflation with graceful exit and reheating} —  
A black hole-dominated early Universe remains a plausible scenario~\cite{Nagatani:1998gv,Conzinu:2023fui}, particularly if PBHs formed via quantum gravity processes~\cite{DeLorenzo:2014pta, Shafiee:2022jfx, Bonanno:2020fgp}. We consider thus an initial population of randomly distributed Hayward PBHs, dominating the pre-BBN Universe's energy density. For analytical tractability, we assume a monochromatic mass spectrum. Interestingly, even within the standard Hot Big Bang (HBB) cosmology, PBHs will eventually dominate due to the slower dilution of their energy density compared to that of radiation~\cite{Hidalgo:2011fj,Suyama:2014vga,Zagorac:2019ekv,Hooper:2019gtx}.

In a spatially flat Universe ($k=0$), the modified Hubble rate~\eqref{eq:H_2_Hayward} will be recast as
\begin{equation}
H^2=\frac{8\pi}{3}G_\mathrm{N}\left( \rho^{-1} +\frac{8\pi}{3}G_\mathrm{N}L^2\right)^{-1}.
\label{modH^2}
\end{equation}
This equation implies a non-singular evolution, as expected for a Universe filled with regular black holes. At high densities, $H^2 \simeq L^{-2}$, and one is met with a nearly exponential cosmic expansion $
a(t) \simeq a_\mathrm{i} e^{t/L},$ where $a_\mathrm{i}$ is the initial scale factor. Inflation ends then when either black holes evaporate or when the energy density drops to $\rho_{\rm e} \simeq G_\mathrm{N}^{-1}L^{-2}$, i.e. when both terms in Eq.~\eqref{modH^2} become comparable. In the post-inflationary era however, if evaporation has not yet been occurred, cosmic expansion will proceed as:
\begin{equation}
H^2 = \frac{8\pi}{3}G_\mathrm{N} \frac{\rho_{\rm e} a^3_{\rm e}}{a^3},
\end{equation}
since in that case $\rho \simeq \rho_{\rm BH} = \rho_{\rm e} a^3_{\rm e} / a^3$, when the index $\mathrm{e}$ stands for the time at the end of inflation. Note that $\rho_{\rm BH}$ is the PBH mass density.  

PBHs will evaporate via Hawking radiation~\cite{Hawking:1974rv}, with a lifetime dependent on their mass. Evaporation may occur either during or after inflation, driving a natural exit from the accelerated cosmic expansion phase and a natural transition to the radiation-dominated era where $H^2 \propto a^{-4}$.

Having discussed above, the evolution of the background dynamics let us discuss here the inflationary predictions of our cosmic acceleration scenario. In particular, the number of inflationary e-folds, $ N_\mathrm{inf} \equiv \ln(a_\mathrm{e}/a_\mathrm{i}) $, must satisfy the horizon problem constraint. Assuming evaporation occurs after inflation ends we obtain that 
\begin{equation}
\rho _{\rm e} \simeq G_\mathrm{N}^{-1} L^{-2} = \frac{\pi ^2}{30} g_{\rm reh} T _{\rm reh} ^4,
\end{equation}
with $g_{\rm reh}$ the relativistic degrees of freedom at reheating. The reheating temperature reads then as
\begin{equation}
T_{\rm reh} = \left( \frac{30}{\pi^2 g_{\rm reh}} \right)^{1/4} L^{-1/2} G_\mathrm{N}^{-1/4}.
\end{equation}
Since $T \propto 1/a$, we have $a_{\rm e}/a_0 \sim T_0/T_{\rm reh}$ with $T_0 = 2.7$ K and $a_0 = 1$.
Requiring thus that the particle horizon during inflation exceeds the today’s observable comoving scale we obtain that
\begin{equation}
\label{eq:N_inf_constraint}
N_\mathrm{inf} > \ln(T_0/T_{\rm reh}) - \ln(L H_0).
\end{equation}
This bound is easily satisfied for $L \gtrsim l_\mathrm{Pl}$ and sufficiently small $a_\mathrm{i}$.

Furthermore, going at the perturbative level, one should comment that since we find a nearly exponential cosmic expansion, the first and second Hubble flow slow-roll parameters $\epsilon_1$ and $\epsilon_2$ defined as $\epsilon_1 \equiv -\dot{H}/H^2$ and $\epsilon_2 \equiv \frac{1}{\epsilon_1}\frac{\mathrm{d}\epsilon_1}{\mathrm{d}N}$ should be very small, leading to a spectral index $n_\mathrm{s} \simeq 1 - 2\epsilon_1 - \epsilon_2 \simeq 1$ and a tensor-to-scalar ratio $r\simeq 16\epsilon_1 \ll 1$ during inflation, compatible with the Planck CMB data~\cite{Planck:2018vyg}.

With regard to reheating, the latter will occur naturally through PBH evaporation. Using the time-dependent Hayward metric~\cite{Frolov:2017rjz}:
\begin{equation}
\label{Hayward}
F = 1 - \frac{2 G_\mathrm{N} m(t) R^2}{R^3 + 2 G_\mathrm{N} m(t) L^2},
\end{equation}
the mass loss rate will be recast as
\begin{equation}
\label{evap}
\frac{dm(t)}{dt} \sim - \frac{1}{C^3 G_\mathrm{N}^2 m(t)^2},
\end{equation}
where $ C = \left(\frac{640\pi}{n_{\rm p}}\right)^{1/3} \left(\frac{L}{l_\mathrm{Pl}}\right)^{2/3} $ and $n_{\rm p}$ is the number of particle polarizations. Integrating thus \Eq{evap} yields the black hole lifetime:
\begin{equation}
\label{firstconstraint}
t_{\rm evap} = \frac{1}{3} C^3 G_\mathrm{N}^2 m^3.
\end{equation}
To ensure also that evaporation completes before BBN ($t_{\rm BBN} \sim 1\, \text{min}$), we impose:
\begin{equation}
\label{constr1}
t_{\rm evap} \leq t_{\rm BBN} \Rightarrow m \leq G_\mathrm{N}^{-2/3} C^{-1} (3 t_{\rm BBN})^{1/3}.
\end{equation}
For $n_{\rm p} \sim 100$ and $L = 100\,l_{\rm Pl}$, we get straightforwardly that $m < 5 \times 10^8$ g.
As long as Eqs.~\eqref{eq:N_inf_constraint} and \eqref{constr1} 
are satisfied, our scenario ensures a consistent inflationary phase with graceful exit and reheating proceeding via PBH evaporation.

To determine now the end of the early cosmic acceleration phase, one needs to compare the critical density $\rho_{\rm c}$ \eqref{rhoch} with the energy density at PBH evaporation time $\rho_{\rm evap}$, assuming a radiation-dominated post-evaporation era. Using thus Eq.~\eqref{firstconstraint} together with the fact that $H_{\rm evap} = 1/2t_{\rm evap}$, we find that
\begin{equation}
\label{eq:rho_evap}
\rho _{\rm evap} = 27648 \pi ^4 \Mp ^4 \left(\frac{\Mp}{m}\right)^6 \left(\frac{n_{\rm p}}{640\pi}\right)^{2/3} \left(\frac{l _{\rm Pl}}{L}\right)^{4/3},
\end{equation}
where $\Mp ^2 = 1 / (8\pi G_\mathrm{N})$.

\begin{figure*}[ht!]
\centering
\includegraphics[width=0.49\textwidth]{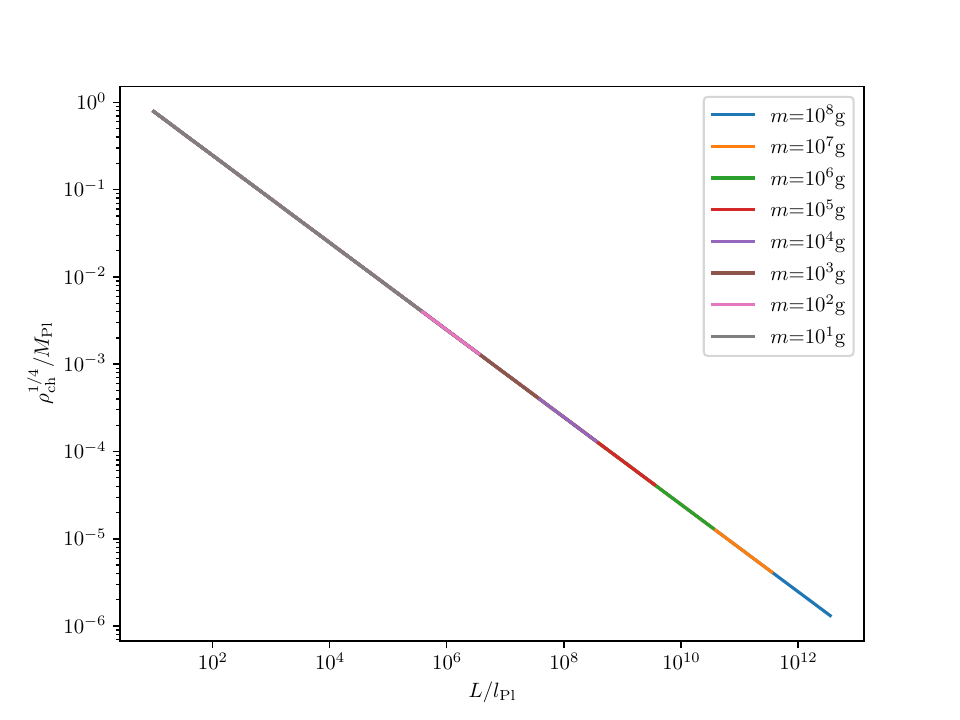}
\includegraphics[width=0.49\textwidth]{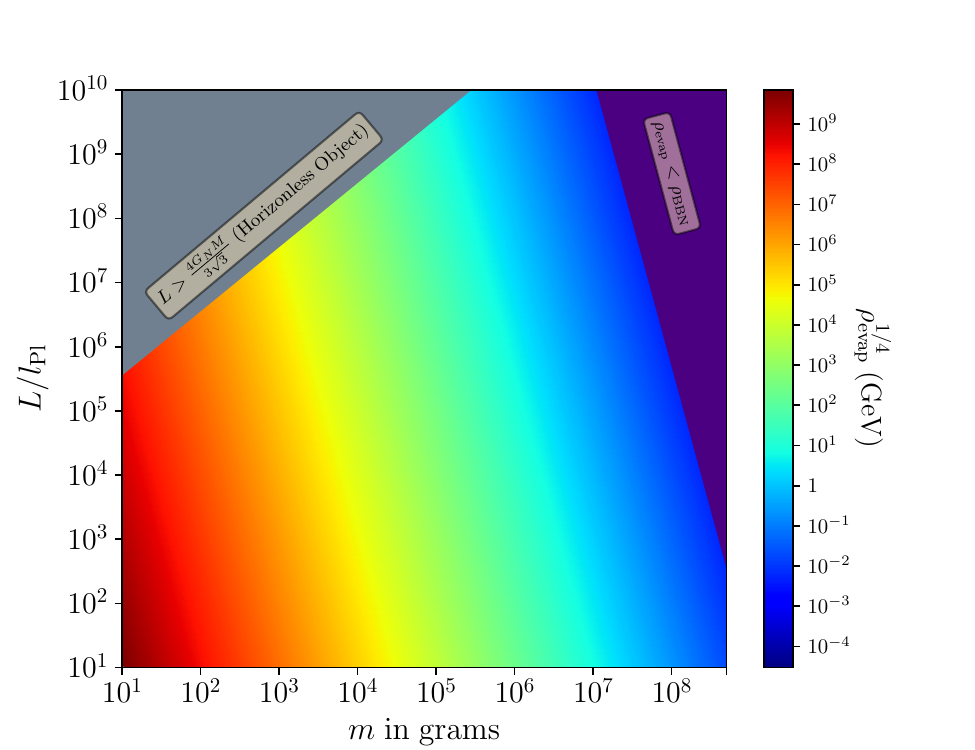}
\caption{{Left Panel: We show $\rho_\mathrm{c}$ as a function of the regularising length scale $L$. $L$ has un upper bound limit reading as $L<\frac{4G_\mathrm{N}m}{3\sqrt{3}}$. Right Panel: We show $\rho_\mathrm{evap}$ (color-bar axis) as a function of the PBH mass $m$ ($x$-axis) and the scale $L$ ($y$-axis). The grey region $L>\frac{4G_\mathrm{N}m}{3\sqrt{3}}$ is not particularly interesting since there we obtain horizonless objects while the magenta one, where $\rho_\mathrm{evap}<\rho_\mathrm{BBN}$ is theoretically excluded.}}
\label{fig:rho_c_vs_rho_evap}
\end{figure*}

In \Fig{fig:rho_c_vs_rho_evap}, the left panel shows $\rho_{\rm c}$ vs. $L$, with the constraint $L < \frac{4G_\mathrm{N}m}{3\sqrt{3}}$ ensuring the presence of a horizon. The right panel displays $\rho_{\rm evap}$ as a function of $m$ and $L$. The grey region (no horizons) and magenta region (evaporation after BBN) are excluded. In particular, the magenta region determined by the condition $\rho_{\rm evap} < \rho_{\rm BBN}$, can be described by
\begin{equation}
L > 7 \times 10^5\, l_\mathrm{Pl} \left( \frac{10\, \text{MeV}}{\rho_\mathrm{BBN}^{1/4}} \right)^3 \left( \frac{n_{\rm p}}{640\pi} \right)^{1/2} \left( \frac{10^8\, \text{g}}{m} \right)^{9/2}.
\end{equation}

As shown from \Fig{fig:rho_c_vs_rho_evap}, $\rho_{\rm c} > \rho_{\rm evap}$ always holds, so cosmic acceleration ends before PBH evaporation. Similar results apply to the Bardeen and Dymnikova spacetimes, where $t_{\rm evap} \propto \tilde{C} t_{\rm S}$ with $\tilde{C} > 1$~\cite{Calza:2024fzo}. For the singular Schwarzschild–de Sitter case, the early cosmic acceleration phase always ends due to PBH evaporation [See \cite{Dialektopoulos:2025mfz} for more details].

{\textit{Early dark energy contribution before matter-radiation equality--}}
Let us now account for an alternative cosmic scenario, where inflation is driven by a scalar field or any other possible cosmological mechanism. Repulsive PBHs can then form due to quantum gravity gravitational collapse processes. ~\cite{1968qtr..conf...87B,Bojowald:2005qw,DeLorenzo:2014pta, Bonanno:2020fgp,Shafiee:2022jfx,Carballo-Rubio:2023mvr,Bambi:2023try,Bonanno:2023rzk,Harada:2025cwd} At the end, one is met with a mixture of radiation and matter in the form of PBHs, with their energy densities scaling respectively as 
$\rho_{\rm rad}=\rho_\mathrm{rad,e}\left(\frac{a_\mathrm{e}}{a}\right)^4$ and
$\rho_{\rm PBH}= \rho_{\mathrm{PBH,e}}\left(\frac{a_\mathrm{e}}{a}\right)^n$,
with $n$ not being necessarily equal to three due to the non-standard equation-of-state (EoS) expected for repulsive PBHs and due to cosmological coupling~\cite{Cadoni:2023lum,Cadoni:2023lqe} and with $\rho_{\rm PBH}$ standing for a theoretically consistent definition for the energy density of a PBH with repulsive behaviour.

Depending on the value of the initial PBH abundance and the parameter $n$, PBHs can dominate or not the energy budget of the Universe. To answer this question, one should consider that the PBH abundance scales as 
\beq\label{eq:Omega_PBH}
\Omega_\mathrm{PBH} \equiv \frac{\rho_\mathrm{PBH}}{\rho_\mathrm{rad}} = \Omega_\mathrm{PBH,f} \left(\frac{a}{a_\mathrm{f}}\right)^{4-n},
\eeq 
where $\Omega_\mathrm{PBH,f}$ is the initial abundance of PBHs when they form. 

In particular, for high enough initial PBH abundances and for $n<4$, PBHs will eventually dominate as can be inferred from \Eq{eq:Omega_PBH}. In this case, the cosmic expansion dynamics will be similar to \Eq{eq:H_2_Hayward} with an early cosmic acceleration period before to recombination, often quoted as early dark energy era. 
Such a PBH-driven EDE epoch before CMB emission with $\Omega_\mathrm{PBH} = \Omega_\mathrm{EDE} > 0.5$ is however excluded since it will significantly affect the CMB peaks position~\cite{Khlopov:1985fch} suppressing also the growth of large scale structures~\cite{Ferreira:1997au,Ferreira:1997hj,Doran:2001rw}.

Nevertheless, if the initial PBH abundance is small, the PBH abundance will increase but not at a level such as that PBHs will dominate, i.e., $\Omega_\mathrm{PBH}<0.5$. In this scenario, one is met with a co-existence of radiation and PBH matter. Such a mixture of different energy density components necessitates the developement of an extended ``Swiss Cheese'' cosmological framework (or a Szekeres-like framework)~\cite{Carrera:2008pi,Celerier:2024dvs}. Astonishingly, if PBH evaporation starts slightly before matter-radiation equality, something which corresponds to PBH masses of around $10^{12}\mathrm{g}$, and with PBH abundances within the range $0.107<\Omega_\mathrm{PBH}<0.5$, one is inevitably met with the correct amount of EDE, compatible with current observational CMB~\cite{Pettorino:2013ia} and large scale structure~\cite{Smith:2020rxx} EDE constraints being recast as 
\beq
\Omega_\mathrm{EDE}(t_\mathrm{LS})<0.015\quad\mathrm{and}\quad 0.015<\Omega_\mathrm{EDE}(t_\mathrm{eq})<0.107,
\eeq
where $t_\mathrm{LS}$ and $t_\mathrm{eq}$ denote respectively the times at the last-scattering and matter-radiation equality. Notably, in such a scenario we have a PBH-EDE component without the need to introduce hypothetical scalar fields as usually adopted in the literature~\cite{Poulin:2023lkg}. An interesting feature of our mechanism is the fact that our PBH-EDE component decays faster than radiation due to Hawking black hole evaporation, a condition which is necessary in order to obtain an increased early value of the Hubble parameter. Our PBH-driven EDE proposed mechanism can then be compatible with the late-cosmology SNIa observations~\cite{Poulin:2018cxd,Kumar:2024soe,Shah:2023sna}, accounting thus in a natural way for the $H_0$ tension.

{\bf{Conclusions}}--We explore in this essay a novel mechanism for cosmic acceleration driven by PBHs with repulsive behavior, within the framework of “Swiss-Cheese” cosmology. Remarkably, matching such PBH black hole spacetimes to a homogeneous and isotropic cosmological background one inevitably finds an early cosmic acceleration era, which terminates naturally via PBH evaporation or below a critical energy scale set by the PBH regularization length scale.

Remarkably, this generic PBH-driven acceleration can account for an almost exponential early inflationary era or an EDE component prior to the emission of CMB. In particular, ultra-light PBHs with masses $ m < 5 \times 10^8\,\mathrm{g} $ can drive inflation with a graceful exit and successful reheating with the latter proceeding through the process of PBH evaporation. Meanwhile, PBHs with $ m \sim 10^{12}\,\mathrm{g} $ and abundances $ 0.107 < \Omega^\mathrm{eq}_\mathrm{PBH} < 0.5 $ just before matter-radiation equality can generate sufficient EDE to ease the Hubble tension.

\begin{figure}[h!]
\centering
\includegraphics[width=0.8\textwidth]{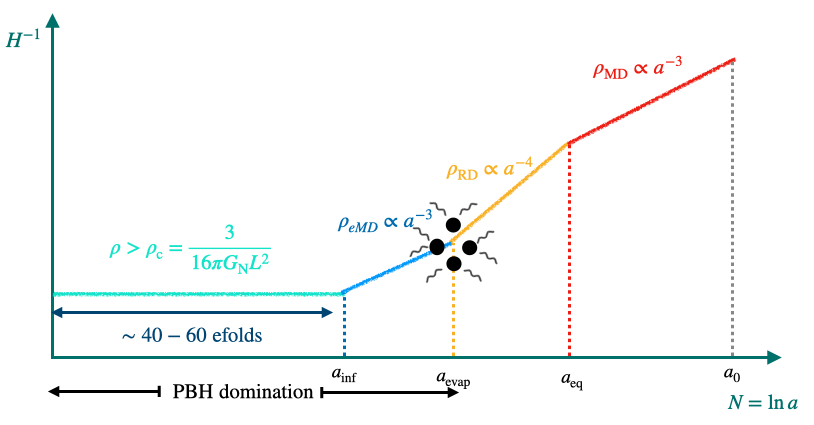}
\caption{{The evolution of the cosmological horizon $H^{-1}$ for a Universe filled with ``repulsive-like" primordial black holes of Hayward type.}}
\label{fig:cosmic_history}
\end{figure}

Our novel cosmological scenario giving rise to early cosmic acceleration is illustrated with  \Fig{fig:cosmic_history} where we schematically display the dynamics of the cosmological horizon scale $H^{-1}$ as a function of the e-fold number $N$ for a Universe filled with Hayward-type PBHs. Similar cosmological horizon dynamics is expected for the Bardeen, Dymnikova and the Schwarzschild-de Sitter black holes as well. 

The new mechanism of cosmic acceleration proposed in \cite{Dialektopoulos:2025mfz} opens several new interesting possibilities that are worth further study, working with specific quantum gravity models. Interestingly, PBH-driven early matter dominated (eMD) eras, like the ones studied here, can lead to early structure formation. In particular, sub-horizon energy density fluctuations during an eMD era grow as $\frac{\delta \rho}{\rho}\propto a$~\cite{Mukhanov:1990me}, facilitating thus naturally structure formation. This possibility has been recently considered in~\cite{Jedamzik:2010dq, Barenboim:2013gya,Eggemeier:2020zeg,Hidalgo:2022yed,Domenech:2023afs} and is also connected with a very rich gravitational wave (GW) phenomenology~\cite{Dalianis:2020gup,Fernandez:2023ddy,Dalianis:2024kjr}. PBH-driven eMD eras can probed as well through GWs induced by PBH number density isocurvature perturbations~\cite{Papanikolaou:2020qtd,Domenech:2020ssp,Papanikolaou:2022chm} as well as through GWs sourced by Hawking-radiated gravitons~\cite{Anantua:2008am,Dong:2015yjs,Ireland:2023avg}.

Finally, regarding the late cosmology era, we need to stress that this mechanism may also have relevance for late-time cosmic acceleration, which may be relevant to the so called dark energy problem. This case was explored in~\cite{Kofinas:2017gfv, Anagnostopoulos:2018jdq} using Schwarzschild-de Sitter black holes within the asymptotic safety-modified “Swiss-Cheese” cosmological framework. However, further improvement is needed using a non homogeneous Szekeres type of models~\cite{Carrera:2008pi,Celerier:2024dvs}.

{\bf{Acknowledgments}}--
The authors acknowledge the contribution of the COST Action CA21136 ``Addressing observational tensions in cosmology with systematics and fundamental physics (CosmoVerse)''. K.F.D. was supported by the PNRR-III-C9-2022–I9 call, with project number 760016/27.01.2023. TP acknowledges the contribution of the LISA Cosmology Working Group and the COST Action ``CA23130 - Bridging high and low energies in search of quantum gravity (BridgeQG)'' as well as the support of the INFN Sezione di Napoli \textit{iniziativa specifica} QGSKY. The author names are shown alphabetically.

\bibliography{references}

\end{document}